\documentstyle[12pt]{article}
\def\k{{\rm {\bf k}}}
\def\p{{\rm {\bf p}}}
\def\q{{\rm {\bf q}}}
\def\r{{\rm {\bf r}}}
\def\n{{\rm {\bf n}}}
\def\u{{\rm {\bf u}}}

\begin{document}

\hfill BI-TP 93/77

\hfill December 1993

\vspace{1.5cm}

\begin{center}
{\bf THE NONPERTURBATIVE EQUATION FOR THE INFRARED
${\bf \Pi_{44}(0)}$-LIMIT IN THE TEMPORAL AXIAL GAUGE}
\footnote{Research is partially supported by "Volkswagen-Stiftung"}

\vspace{1.5cm}
{\bf O.K.Kalashnikov}
\footnote{Permanent address: Department of
Theoretical Physics, P.N.Lebedev Physical Institute, Russian
Academy of Sciences, 117924 Moscow, Russia. E-mail address:
kalash@td.fian.free.msk.su}

Fakult\"at f\"ur Physik

Universit\"at Bielefeld

D-33501 Bielefeld, Germany

\vspace{2.5cm}
{\bf Abstract}
\end{center}

The nonperturbative equation for the infrared $\Pi_{44}(0)$-limit
is built by using the Slavnov-Taylor identity to define
the three-gluon vertex function in the temporal axial gauge.
We found that all vertex corrections should be taken into
account along with the standard ring graphs to keep the gauge
covariance throughout calculations and to give correctly the
nonperturbative $g^3$-term. This term is explicitly calculated and
compared with the previously known results.

\newpage

Recently the interest has been revived (see e.g. Refs.[1,2])
to calculate of the Debye mass beyond the leading term
and to investigate of its gauge dependence. This problem
is strongly connected with the analogous calculations of the infrared
$\Pi_{44}(0)$-limit which are more simple but today they are
reliably known only up to $g^2$-order. The next-to-leading term
(the one of order $g^3$) was found many years ago (at first in paper
[3] and then in other papers [4,5] and [6]) but till now its accuracy
is not confirmed. Unfortunately these results are qualitatively
different and show that the $g^3$-term found for the infrared
$\Pi_{44}$-limit is a gauge-dependent quantity and its coefficient
(that is more essential) is very sensitive to keeping the gauge
covariance throughout calculations. The problem aggravates when
the $\Pi_{44}(0,\p)$-quantity is calculated but namely this limit
(as it was shown in Ref.[1,2]) is needed to define the Debye screening.
However there are many reasons at first to find reliably the infrared
$\Pi_{44}(0)$-limit by using the temporal axial gauge since this gauge
is singled out both for keeping the gauge covariance and for calculating
the Debye screening.

The goal of this paper is to derive the nonperturbative equation for
the infrared $\Pi_{44}(0)$-limit by operating the standard Green
function technique within the temporal axial gauge. The obtained equation
takes into account all perturbative graphs (the ring graphs as well as
the vertex corrections) and its gauge covariance is guaranteed by
exploiting the exact Slavnov-Taylor identities to find the
nonperturbative three-gluon vertex. This vertex is qualitatively
different from the bare one and the derived equation is free from any
divergencies. The $g^3$-term is found to be a positive correction to
the leading one and we compare it with the previously known results.

It is well-known that the temporal axial gauge is convenient
for building the nonperturbative schemes since the choice of
the gauge vector $\n_\mu$  to be parallel to the medium one $\u_\mu$
considerably simplifies the Green function
technique. The exact polarization tensor (in the axial gauge) is
determined by only two tensor structures [6]
$$
\Pi_{\mu\nu}(k)=G\left(\delta_{\mu\nu}-\frac{k_\mu k_\nu}{k^2}\right)+
(F-G)B_{\mu\nu}\,,\
\eqno{(1)}
$$
and the gluon propagator has a rather simple form
$$
{\cal D}_{ij}(k)=
\frac{1}{k^2+G}\left(\delta_{ij}-\frac{k_ik_j}{\k^2}\right)+
\frac{1}{k^2+F}\frac{k^2}{k_4^2}\frac{k_ik_j}{\k^2}.
\eqno{(2)}
$$
The scalar functions $F(k)$ and $G(k)$ are defined as
follows
$$
G(k)=\frac{1}{2}\left(\sum_{i}\Pi_{ii}(k)+
\frac{k_4^2}{\k^2}\Pi_{44}(k)\right)\,,\qquad
F(k)=\frac{k^2}{\k^2}\Pi_{44}(k)\,,\
\eqno{(3)}
$$
and they should be calculated through the graph (or another)
representation for $\Pi$. Due to a peculiarity of the temporal
axial gauge the functions ${\cal D}_{44}(k)$ and ${\cal D}_{4j}(k)$
are completely eliminated from the formalism but there is a very specific
singularity $(k_4^2=0)$ which requires a very delicate treatment.
The exact Slavnov-Taylor identity for the three-gluon vertex
function has a rather simple form
$$
r_\mu\Gamma_{\mu\nu\gamma}^{abc}(r,\,p,\,q)=
igf^{abc}[{\cal D}_{\nu\gamma}^{-1}(p)-{\cal D}_{\nu\gamma}^{-1}(q)]\,,\
\eqno{(4)}
$$
and namely this fact is a doubtless advantage of the axial gauge.
Eq.(4) is our main instrument and we exploit it to build the
nonperturbative vertex function for calculating the infrared
$\Pi_{44}(0)$-limit.
We also use Eq.(4) in its differential form which allows in many
cases (see e.g. Ref.[7]) to define the exact infrared limit of the
three-gluon vertex in a very convenient manner. For example, the
infrared $\Gamma_{4ij}^{abc}(-p,0,p)$-limit can be easily found
from the standard identity
$$
\Gamma_{4ij}^{abc}(-p,\,0,\,p)=-
igf^{abc}\frac{\partial {\cal D}_{4j}^{-1}(p)}{\partial p_i}\,,\
\eqno{(5)}
$$
which directly results from Eq.(4). This limit being exact has a
rather simple form
$$
\Gamma_{4ij}^{abc}(-p,\,0,\,p)=igf^{abc}
\left\{\delta_{ij}[1+\frac{F(p)}{p^2}]+
p_j\frac{\partial}{\partial p_i}
\left[\frac{F(p)}{p^2}\right]\right\}p_4\,,\
\eqno{(6)}
$$
and depends on one function which determines
the usual representation for the ${\cal D}_{4i}^{-1}(p)$-propagator
$$
{\cal D}_{4j}^{-1}(p)=-\left(1+\frac{F(p)}{p^2}\right)p_jp_4 .
\eqno{(7)}
$$
Unfortunately the limit (6) is not our case but
we shall return back to Eq.(6) when the nonperturbative expression
for the infrared $\Gamma_{4ij}^{abc}(0,p,-p)$-limit is discussed.

The exact graph representation for the gluon polarization tensor
is well-known (see e.g. papers [6,8]) and contains (in the axial gauge)
the standard four nonperturbative graphs. However if one considers
the $\Pi_{44}$-components only two one-loop nonperturbative graphs
are essential since the rest graphs (the two very complicated ones)
are equal to zero exactly. The analytical expression for the first two
graphs has a rather simple form and after some algebra being performed
(by taking into account that $\Gamma_{ij4}^{abc}=-igf^{abc}\Gamma_{ij4}$)
the equation for $m_{E}^2$ (where
$m_{E}^2=\Pi_{44}(0)$) is found to be
\setcounter{equation}{7}
\begin{eqnarray}
m_E^2&=&
\frac{g^2N}{\beta}\sum_{p_4}
\int\frac{d^3\p}{(2\pi)^3}{\cal D}_{ii}(p)\\
&-&\frac{g^2N}{2\beta}\sum_{p_4}
\int\frac{d^3\p}{(2\pi)^3}
2p_{4}\left[{\cal D}_{li}(p)
\Gamma_{ij4}(p,\, -p,\, 0){\cal D}_{jl}(p)\right]\,,\
\nonumber\end{eqnarray}
and we are going to solve it keeping the gauge covariance at each
step of the calculations. Here all functions (including the vertex one)
are exact and our main problem is to find the nonperturbative
expression for the infrared $\Gamma_{ij4}(p,-p,0)$-limit.

Unfortunately the exact expression for the infrared
$\Gamma_{ij4}(p,-p,0)$-limit (which is not Eq.(6)) lies beyond our
possibilities and therefore only its nonperturbative ansatz will be
represented by following the more general formula obtained in Ref.[7].
This formula is found to be
\setcounter{equation}{8}
\begin{eqnarray}
&&\Gamma_{4ij}^{abc}(q,\,r,\,p)=-igf^{abc}\left\{\delta_{ij}
(r_4-p_4)-\frac{1}{r^2-p^2}\left[\left(\frac{G(r)}{\r^2}-
\frac{F(r)}{r^2}\frac{r_4^2}{\r^2}\right)\right.\right.\nonumber\\
&&\hspace{2em}
-\left.\left.\left(\frac{G(p)}{\p^2}-\frac{F(p)}{p^2}\frac{p_4^2}
{\p^2}\right)\right][(\p\r)\delta_{ij}-p_ir_j](r_4-p_4)
\right.\nonumber\\
&&\hspace{2em}
+\left.\delta_{ij}\left(r_4\frac{F(r)}{r^2}-\frac{F(p)}{p^2}p_4\right)
+\frac{1}{q^2-r^2}\left(\frac{F(q)}{q^2}-\frac{F(r)}{r^2}\right)
q_ir_4(q-r)_j\right.\nonumber\\
&&\hspace{2em}
\left.+\frac{1}{p^2-q^2}\left(\frac{F(p)}{p^2}
-\frac{F(q)}{q^2}\right)(p-q)_ip_4q_j\right.\nonumber\\
&&\hspace{2em}
\left.-\frac{1}{r^2-p^2}\left(\frac{F(r)}{r^2}-\frac{F(p)}{p^2}\right)
r_4p_4(r-p)_4\delta_{ij}\right\}\,,\
\end{eqnarray}
and it is valid for any momentum set including the soft domain.
To find Eq.(9) the standard inverse gluon propagator is used
$$
{\cal {D}}_{ij}^{-1}(p)=\left(\delta_{il}-\frac{p_{i}p_{j}}{\p^2}
\right)(p^2+G(p))+\left(1+\frac{F(p)}{p^2}\right)p_{4}^2
\frac{p_{i}p_{j}}{\p^2}\,,\
\eqno{(10)}
$$
and the exact Slavnov-Taylor identities were exploited (see Ref.[7]
for details). The transversal part of the $\Gamma_{4ij}^{abc}(q,r,p)$-
function is omitted from Eq.(9) since it is not essential for what
follows. Of course, it is necessary to bear in mind to exclude all
singularities from Eq.(9) for any momentum going to zero.

The vertex function (9) easily reproduces
the exact formula (6) if one momentum goes to zero (at first $r_4=0$
and then $|\r|\rightarrow 0$) but it is more essential that one can
exploit this representation in a more general case to find the
infrared $\Gamma_{ij4}^{abc}(p,-p,0)$-limit. The final result has the form
\setcounter{equation}{10}
\begin{eqnarray}
&&\Gamma_{ij4}^{abc}(p,\,-p,\,0)=-igf^{abc}\left\{2\delta_{ij}
\left(1+\frac{F(p)}{p^2}\right)\right.\nonumber\\
&&\hspace{2em}
+\left.2\frac{p_4^2}{\p^2}\left(\frac{F(p)}{p^2}\right)
\left(\delta_{ij}-\frac{p_{i}p_{j}}{\p^2} \right)
+\frac{1}{|\p|}\left[\frac{\partial}{\partial |\p|}\left(\frac{G(p)}
{\p^2}\right)\right][\p^2\delta_{ij}-p_ip_j]\right.\nonumber\\
&&\hspace{2em}
+\left.p_4^2\left[\frac{1}{|\p|}\frac{\partial}
{\partial |\p|}
\left(\frac{F(p)}{p^2}\right)\right]\frac{p_ip_j}{\p^2}\right\}p_4\,,\
\end{eqnarray}
and we apply this representation for treating Eq.(8). The vertex found
is qualitatevly different from the bare one  and we hope that Eq.(11)
will be usefull for many applications.

All algebra within Eq.(8) is very simple and therefore omitted. The
equation for $m_E^2$ is found as follows
\begin {eqnarray}
&&m_E^2=-\frac{g^2N}{\beta}\sum_{p_{4}}\int\frac{d^3\p}{(2\pi)^3}
\left\{
\frac{1}{p_4^2}\frac{1}{1+\frac{F(p)}{p^2}}-
\frac{2}{p^2+G(p)}\right.\nonumber\\
&&\hspace{2em}\left.
+\frac{4p_4^2}{[p^2+G(p)]^2}
+\frac{1}{|\p|}\left[\frac{\partial}{\partial |\p|}
\left(\frac{F(p)}{p^2}\right)\right]
\left[1+\frac{F(p)}{p^2}\right]^{-2}
\right.\nonumber\\
&&\hspace{2em}\left.+
\left[\frac{2p_4^2}{|\p|}\left(\frac{\partial G(p)}{\partial |p|}\right)
-4p_4^2\left(\frac{G(p)}{\p^2}-\frac{F(p)}{\p^2}\right)\right]\frac{1}
{[p^2+G(p)]^2}
\right\}\,,\
\end{eqnarray}
and it is the main subject of the following discussion. Eq.(12)
correctly summarizes not only the ring graphs but essentially exploits
the dressed three-gluon vertex in a nonperturbative manner and it is
very probably that this equation is exact for the $g^3$-term. Of course,
Eq.(12) correctly reproduces the leading order for $m_E^2$ if the
functions G(p) and F(p) are omitted
$$
m_{E}^2=\frac{g^2N}{\beta}\sum_{p_4}\int\frac{d^3\p}{(2\pi)^3}
\frac{\partial}{\partial p_{4}}\left[\frac{2p_4}{p^2}+
\frac{1}{p_4}\right]\,,\
\eqno{(13)}
$$
and the standard regularization for the temporal axial gauge is used
$$
\frac{1}{\beta}\sum_{p_4}\frac{1}{p_4^2}=0 .
\eqno{(14)}
$$
No other terms should be taken into account in Eq.(12) since the
rest graphs (which usually determine the $\Pi_{\mu\nu}$-tensor) are
exactly equal to zero if the $\Pi_{44}$-quantity is only considered.
This fact is due to a simple Lorentz tensor structure of the bare
$\Gamma_{4}$-vertex function and the specific feature of the temporal
axial gauge where the ${\cal D}_{44}$-function is eliminated from the
formalism. The $g^3$-term (as well as all the other ones) should be
completely determined by Eq.(12) and we found that this equation is
free from any divergencies (of course if the full sum over $p_4$ is
taken into account).

Our next task is to find the appropriate equation for calculating the
$g^3$-term on the basis of Eq.(12). This is a pure nonperturbative term
and arises within Eq.(12) when the soft momentum region is exploited.
Due to the different infrared behaviour of the functions G(p) and F(p)
only the latter gives the appropriate contribution to reproduce the
$g^3$-term and all other terms being of $g^4$-order can be omitted. The
final equation for the $g^3$-term (here the $\delta m_E^2$-term)
has a rather simple form
\setcounter{equation}{14}
\begin{eqnarray}
\delta m_E^2&=&-\frac{g^2N}{\beta}\sum_{p_4}\int\frac{d^3\p}{(2\pi)^3}
\left\{\frac{1}{p_4^2}\frac{1}{1+\frac{F(p)}{p^2}}\right.\nonumber\\
&&+\left.\frac{1}{|\p|}\left[\frac{\partial}{\partial |\p|}
\left(\frac{F(p)}{p^2}\right)\right]\left[1+\frac{F(p)}{p^2}
\right]^{-2}\right\}\,,\
\end{eqnarray}
and it can be solved independently from Eq.(12). However,
there is a question with the first term in Eq.(15) which contains the
specific singularity of the temporal axial gauge and its analytical
behaviour is not clear. Nevertheless we insist that this term is equal
to zero if only the static $\Pi_{44}(0)$-limit is used
$$
\Pi_{44}^{(2)}(p_4=0,|\p|\rightarrow 0)=\frac{g^2N}{3\beta^2}\,,\
\eqno{(16)}
$$
and all calculations are performed in the standard infrared manner
(when the sum over $p_4$ is replaced by one term with $p_4=0$). Finally
the above equation is found to be
\setcounter{equation}{16}
\begin{eqnarray}
\delta m_E^2=-\frac{g^2N}{\beta}\int\frac{d^3\p}{(2\pi)^3}
\frac{1}{|\p|}\left[\frac{\partial}{\partial |\p|}\left(
\frac{\Pi_{44}^{(2)}(0)}{\p^2}\right)\right]
\left[1+\frac{\Pi_{44}^{(2)}(0)}{\p^2}\right]^{-2}\,,\
\end{eqnarray}
where all functions are known and the integral is calculated in the
usual manner. Our result has a rather simple form
$$
m_E^2=[\frac{g^2N}{3}+\frac{3}{4\pi}(\frac{g^2N}{3})^{3/2}]T^2\,,\
\eqno{(18)}
$$
and it is in an agreement with the results [1,2] and [5] if those are
considered in the Feyman gauge only. The other results (see Ref.[3,4])
should be checked to solve reliably the question of the gauge
dependence of the $g^3$-term.

In conclusion we note that the result found for the infrared
$\Pi_{44}(0)$-limit is not reproduced through the effective action
calculated in a constant background field ( see e.g. Refs.[9,10]).
This is the case only for the leading term but in a general case
the correspondence seems to be more complicated. This fact needs
further investigations to establish finally the status of the $g^3$-term
and its connection with the Debye screening. However there is a hope
that the latter problem can be solved independently from the scenario
with the magnetic mass and the question with the magnetic screening
is separated (at least in the temporal axial gauge) into the
outstanding problem. There is also a nonperturbative equation [11]
\setcounter{equation}{18}
\begin{eqnarray}
m_{M}^2&=&\frac{3N^2g^4}{4\beta^2}\sum_{p_4,q_4,r_4}
(2\pi)^3\int\frac{d^3\p}{(2\pi)^3}\frac{d^3\q}{(2\pi)^3}
\frac{d^3\r}{(2\pi)^3}\delta^{(4)}(p+q+r)\times\nonumber\\
&\times&{\cal D}_{jn}(q){\cal D}_{it}(r)\frac{\partial}{\partial p_i}
\left[{\cal D}_{nm}(p)\Gamma_{mjt}(-p,-q,-r)\right]
\end{eqnarray}
but its solution which encounters the infrared divergencies
is still unconfirmed. It is important that Eq.(19) is
also accessible only in the temporal axial gauge and it results from
the two rest nonperturbative graphs which are equal to zero in the
above case.

\newpage

\begin{center}
{\bf Acknowledgements}
\end{center}

I would like to thank Rudolf Baier for useful discussions and
all the colleagues from the Department of Theoretical Physics of the
Bielefeld University for the kind hospitality.

\vspace {1.0cm}

\begin{center}
{\bf References}
\end{center}

1. A.K.Rebhan, Phys. Rev. {\bf D48} (1993) R3967.

2. A.K.Rebhan, preprint BI-TP 93/52 (hep-ph/9310202).

3. K.Kajantie and J. Kapusta, Phys. Lett. {\bf 110B} (1982) 299.

4. T.Furusawa and K.Kikkawa, Phys. Lett. {\bf 128B} (1983) 218.

5. T.Toimela, Z. Phys. {\bf C27} (1985) 289.

6. K.Kajantie and J.Kapusta, Ann. Phys. (N.Y.) {\bf 160} (1985) 477.

7. O.K.Kalashnikov, JETP Lett. {\bf 39} (1984) 405.

8. O.K.Kalashnikov, Fortschr. Phys. {\bf 32} (1984) 525.

9. K.Enqvist and K.Kajantie, Z.Phys. {\bf C47} (1990) 291.

10. O.K.Kalashnikov, Phys. Lett. {\bf B304} (1993) 453; JETP Lett.

\hspace {1em}  {\bf 57} (1993) 773.

11. O.K.Kalashnikov, JETP Lett. {\bf 41} (1985) 149.

\end{document}